\newcommand{\beq}{\begin{quote}}
\newcommand{\enq}{\end{quote}}
\newcommand{\be}{\begin{equation}}
\newcommand{\en}{\end{equation}}
\newcommand{\del}{\delta}
\newcommand{\om}{\omega}
\newcommand{\Om}{\Omega}

\documentstyle[prb,aps,epsfig,floats]{revtex}
\begin{document}
\draft
\twocolumn[\hsize\textwidth\columnwidth\hsize\csname@twocolumnfalse%
\endcsname

\title{ Periodic  orbits for three particles
with finite angular momentum} 
\date{Sept. 2001}
\author{Michael Nauenberg\\
Department of Physics\\
University of California, Santa Cruz, CA 95064 
}
\maketitle  
\begin{abstract}
New solutions are obtained for the planar  motion of  three equal mass particles
on a common periodic orbit with finite total angular momentum, 
under the action of  attractive  pairwise forces of the form
$1/r^{p+1}$.  It is shown that for $-2 < p \le  0$, Lagrange's 
1772 circular solution is the limiting case of a complex symmetric orbit. 
The evolution of this orbit and  another recently discovered one  
in the shape of a figure eight is investigated  for a range of angular momenta
by a novel numerical method.  Extensions to  n equal mass particles 
and to three particles of different masses are also  discussed briefly. 
\end{abstract}

keywords: nonlinear science, orbital dynamics
\vskip 0.2in
]
\subsection*{Introduction}

Recently C. Moore \cite{moore} applied the  theory of braids 
to find a 
classification of periodic orbits for the motion of several   
particles moving in a plane under the action of pairwise 
attractive power-law forces, obtaining  numerical results 
for the case of 2 and 3 strands. 
In particular, for three particles of
equal mass, he found a solution where these particles
travel on a common orbit in the shape of
a figure eight. This orbit was  found also independently
by  A. Chenciner and R. Montgomery \cite{mont1},\cite{mont2}
who have proved its existence
for the case of zero total angular momentum and inverse square forces.
Moore's numerical result that this orbit  is stable  has also  been
confirmed by computations of C Sim\'o \cite {simo}. 
Previously, one of the simplest known periodic 
orbits without collisions for this problem \cite{newton1}
was Lagrange's classic  solution 
\cite{lagrange} where the  three  particles  form 
the vertices of an equilateral triangle with each particle
moving  on an elliptic orbit with  focus
in the common  center of mass. 
In this paper we show that for attractive power-law forces of the form 
$1/r^{p+1}$  with  $-2 < p \le 0$, the circular
Lagrange  orbit is the limit of a family 
of orbits which are periodic in a uniformly rotating frame of reference.
For small deviations, we find  analytically that this orbit is a distorted
ellipse which, as the angular momentum decreases, 
resembles a distorted  figure eight  shape, but 
it does not approach  the recently discovered  
zero-angular-momentum figure eight orbit 
which has a different winding number. 
For this figure eight orbit, which  exists for
$-2 < p $, we obtain an analytic approximation, 
and we prove by perturbation theory that it
can be extended to finite angular momentum. We also investigate the evolution 
of these two  orbits for a finite range of total angular momentum
by a novel iterative method, and consider  briefly the
case of such periodic orbits for  n equal mass particles, and 
for three particles of different masses.

\subsection*{ Computational method for periodic orbits}

Our method is based on expanding in a Fourier series the Cartesian coordinates 
for the periodic  orbit on which all the three equal mass particles 
travel, and then finding the extremun of the action for this system  
with respect to the 
Fourier coefficients  in a frame rotating with a fixed frequency $\Om$. 
We obtain analytic approximations by perturbation theory, 
and  numerical solutions
by a novel iterative method which  converges rapidly. 
Setting $\theta=\om t$, where  $t$ is the time, $\om=2\pi/\tau$ is 
the frequency, and 
$\tau$ is the period of the orbit), we have

\be
\label{x1}
x(\theta)=\sum_{k=1}^{k=n}a_s(k_o)sin(k_o \theta)+a_c(k_e)cos(k_e \theta)
\en
and
\be
\label{y1}
y(\theta)=\sum_{k=1}^{k=n}b_s(k_e)sin(k_e \theta)+b_c(k_o)cos(k_o\theta),
\en
where $k_o=2k-1$ and $k_e=2k$, excluding multiples of the integer $3$. 
In practice, the number of terms  
$n$ in the Fourier sums,  Eqs. \ref{x1} and  \ref{y1}, 
depends on the desired accuracy of the solution.
The Cartesian  coordinates of each of the three  particles is then given by 
$x_i=x(\theta_i)$ and $y_i=y(\theta_i)$, for $i=1,2,3$,  
and the invariance of the action under the cyclic  transformation 
$1\rightarrow  2 \rightarrow 3 \rightarrow 1$ is satisfied by the requirement  
$\theta_1=\theta$, $\theta_2=\theta+2\pi/3$
and $\theta_3=\theta+4\pi/3$. 
These conditions are satisfied 
by the equations which determine the extremun of  the action    
in a frame rotating with frequency $\Om$ in the plane 
of the orbit \cite{plane} which takes the form 
\be
\label{action1}
A=\int_0^{2\pi}d\theta [K(\theta)-P(\theta)+
\frac{1}{2} \Om^2 I(\theta)+\Om L(\theta)],
\en
where  $K(\theta)$ is the kinetic energy, $P(\theta)$  the potential energy,
$I(\theta)$ the moment of inertia, and $L(\theta)$  the angular momentum
in the rotating frame. These 
are given by the following expressions:
\be
K(\theta)=\frac{1}{2}\om^2 \sum_{i=1}^{i=3}(\frac{dx_i}{d\theta})^2+
(\frac{dy_i}{d\theta})^2, 
\en
\be
\label{potential1}
P(\theta)=-(\frac{1}{p})(\frac{1}{r_{12}^p}+\frac{1}{r_{13}^p}+\frac{1}{r_{23}^p}),  
\en
\be
\label{iner1}
I(\theta)=\sum_{i=1}^{i=3} x_i^2+y_i^2,
\en
and
\be
\label{angm}
L(\theta)=\om \sum_{i=1}^{i=3} x_i \frac{dy_i}{d\theta}-y_i \frac{dx_i}{d\theta},
\en
where $r_{ij}=\sqrt{(x_i-x_j)^2+(y_i-y_j)^2}$ is the
distance between particles  $i$ and $j$, and 
$p$ is  the  power of the force law,

The extrema of the action, Eq. \ref{action1}, are obtained by finding 
the Fourier coefficients for which
the partial derivatives of the action with respect to these 
coefficients vanish, 
\be
\label{ao}
\frac{\partial A}{\partial a_s(k_o)} = u(k_0)a_s(k_o)-f_s(k_o)- v(k_0)b_c(k_o)=0,
\en

\be
\label{bo}
\frac{\partial A}{\partial b_c(k_o)} = u(k_0)b_c(k_o)-g_c(k_o)- v(k_0)a_s(k_o)=0,
\en

\be
\label{be}
\frac{\partial A}{\partial b_s(k_e)} = u(k_0)b_s(k_e)-g_s(k_e)+ v(k_0)a_c(k_e)=0,
\en

\be
\label{ae}
\frac{\partial A}{\partial a_c(k_e)} = u(k_0)a_c(k_e)-f_c(k_e)+ v(k_0)b_s(k_e)=0,
\en

where $u(k)=k^2\om^2+\Om^2$ and  $v(k)=2k\om \Om $.  Here the functions
\be
\label{fso}
f_s(k_o)=\frac{1}{\pi}\int_0^{2\pi}d\theta 
\frac{\partial P}{\partial x_1}sin(k_o\theta),
\en
\be
\label{fce}
f_c(k_e)=\frac{1}{\pi}\int_0^{2\pi}d\theta 
\frac{\partial P}{\partial x_1}cos(k_e\theta),
\en
\be
\label{gse}
g_s(k_e)=\frac{1}{\pi}\int_0^{2\pi}d\theta 
\frac{\partial P}{\partial y_1}sin(k_e\theta),
\en
\be
\label{gco}
g_c(k_o)=\frac{1}{\pi}\int_0^{2\pi}d\theta 
\frac{\partial P}{\partial y_1}cos(k_o\theta).
\en
depend also  on all the Fourier coefficients.

We obtain numerical solutions of these equations by starting with  
the Fourier coefficients to some 
conjectured approximate  period orbit, e.g. 
according to Moore's classification \cite{moore},
and then  evaluating the  partial derivatives
of the action, Eqs. \ref{ao}-\ref{ae}. If the extrema of the action is at 
a maximum (minimum), an improved  orbit is then  obtained by changing 
each  coefficient in  proportion to  plus(minus)
the corresponding partial derivative of the action.
Thus, starting with some 
initial value for the coefficient $a_s(k_o)$, we obtain a new value $a_s(k_o)$ 
by the relation 
\be
a'_s(k_o)=a_s(k_o)+\frac{\del s}{u(k_o)} \frac{\partial A}{\partial a_s(k_o)}
\en
with similar expressions for the other Fourier coefficients   
where $\del s$ is a positive (negative) parameter if the action is a maximum (minimum). 
This procedure is then iterated until the partial derivatives
of the action are sufficiently small to any  desired accuracy.
Our method differs in an essential way from the standard method of steepest descent 
by the important  factor $1/u(k)$ which is  essential  for  
the convergence of the iterations
with moderate values of $\del s$ (order $.1$).

To obtain  analytic approximations of Eqs. \ref{ao}-\ref{ae}, it is
convenient to determine  the  frequencies $\om$ and $\Om$  by 
fixing  the values of $a_s(1)$ and $b_c(1)$, which
give the scale of the orbit and the magnitude of
the total  angular momentum 
in the rest frame, $L_t= L +\Om I$, where  
$L$ is the mean angular momentum  in the rotating frame, Eq. \ref{angm},
\be
\label{ang2}
L=3\om \sum_{k=1}^{k=n}-k_o a_s(k_o) b_c(k_o)+k_e a_c(k_e) b_s(k_e),
\en
and $I$ is the mean moment of inertia, Eq. \ref{iner1},
\be
\label{iner2}
I=\frac{3}{2}\sum_{k=1}^{k=n}a_s(k_o)^2+a_c(k_e)^2+b_s(k_e)^2
+b_c(k_o)^2.
\en
We have
\be
\label{om1}
\om=\frac{1}{2}(\sqrt{\alpha}+\sqrt{\beta})
\en
and
\be
\Om=\frac{1}{2}(\sqrt{\alpha}-\sqrt{\beta}),
\en
where
\be
\alpha=\frac{f_s(1)-g_c(1)}{(1-b_c(1))}
\en
and
\be
\beta =\frac{f_s(1)+g_c(1)}{(1+b_c(1))}
\en
are obtained from Eqs. \ref{ao} and \ref{bo} for $k_o=1$  and $a_s(1)=1$.
These frequencies scale with the size  $s$ of the orbit  according to the
generalized Kepler relations that $\om^2 s^{p+2}$ and $\Om^2 s^{p+2}$
are constants.

The generalization of Lagrange's solution \cite{lagrange} 
is obtained by setting 
the Fourier coefficients $a_c(k_e)$ and $b_s(k_e)$ equal to zero,
which according to Eqs. \ref{ae} and \ref{be} implies  that $f_c(k_e)$ 
and $g_s(k_e)$ also vanish. Small deviations from
Lagrange's  circular solution can be
obtained by setting $a_s(1)=1$ and  $b_c(1)=1-\del$, 
and expanding Eqs. \ref{ao} and \ref{bo} 
in a power  series $\del$. In the limit of vanishing $\del$
\be
\alpha =\frac{1}{3^{p/4}}\sqrt{-p/2}
\en
which shows that these solutions exist only for $p \le 0$.
In particular, for the special case that  $p=-2$ 
corresponding to harmonic forces, $\om=\sqrt{3}$,
$\Om=0$, and all the Fourier coefficients $a_s(k_0)$ and $b_c(k_0)$
vanish for $k_0>1$. 

\subsection* {Numerical solutions and analytic approximations}

Two numerical solutions of the extended Lagrange problem are shown in 
Figs. 1 and 2,
where segments of the orbit occupied by each particle during a third 
of a period are indicated by a dark line, a dotted line
and a dashed line respectively, and ten successive locations 
on these segments are marked at equal intervals of time to 
illustrate  the relative position  and variable speed of the 
three particles moving on  this  orbit. 
In this  solution  
the angular momentum of the orbit decreases  
as $b_c(1)$ decreases,  
and the approximate elliptical orbit 
starts to  get pinched in the middle. 
This is shown in Fig.1 for $p=-.05$  
at the special value $b_c(1)=.049$ 
where the orbit also  develops cusps at the two ends. 
At these values of the parameters each particle comes momentarily to rest 
at these points  because
the attractive forces due to the other two particles balance  
the centrifugal force in the rotating frame. 

\begin{figure}
\begin{center}
\epsfxsize=\columnwidth
\epsfbox{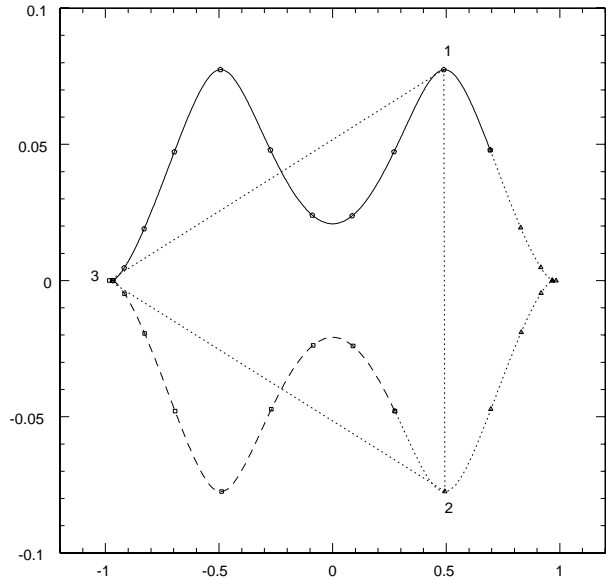}
\end{center}
\caption{
An extension of Lagrange's circular orbit for three equal
mass particles labeled here 1,2 and 3 , showing their location
at ten equal time intervals during one third of a
period. For example, when particle 3 reaches an extremun
along the x-coordinate, its y-component
of acceleration  vanishes by symmetry, and therefore the
location of the other two particles, 1 and 2, lie on the vertices of an
isosceles triangle as shown here for the case $p=-.05$ and $b_c(1)=.049$.
}
\label{equi3}
\end{figure}

Further decrease of  the value of $b_c(1)$, corresponding
to decreasing  angular momentum,  requires that the y-component of velocity
at these ends change  direction, which is possible only  
if an additional lobe appears at each end.  
A further decrease of $b_c(1)$ leads to an additional lobe at the center,
as shown in Fig. 2 for $p=-.05$ and $b_c(1)=.0171 $.  

\begin{figure}
\begin{center}
\epsfxsize=\columnwidth
\epsfbox{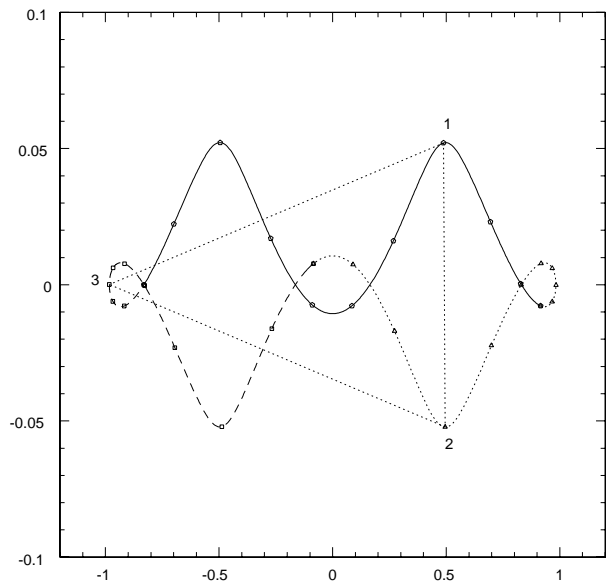}
\end{center}
\caption{ Extended Lagrange orbit with  crossings near the center
and additional lobes at the two ends, for $p=-.05$ and $b_c(1)=.0171$
}
\label{equi2}
\end{figure}

The recently discovered \cite{moore} - \cite{mont2}
zero angular momentum periodic orbit in the shape of a  
figure eight is a solution of  Eqs. \ref{ao} and \ref{be} 
with  the Fourier coefficients $a_c(k_e)$ and $b_c(k_o)$ set equal to zero. 
According to Eqs. \ref{ae} and \ref{bo}, this implies 
that $f_c(k_e)=g_c(k_o)=0$. 
For the special case that $p=2$, it can be shown that the moment of inertia, 
Eq. \ref{iner1}, is a constant. Consequently
the shape of this orbit can be determined from purely geometrical considerations
by imposing  the condition that the Fourier coefficients of an expansion
of  this moment and that of the angular momentum vanish.
Keeping terms in the sums in Eqs. \ref{x1} and  \ref{y1} up to n=4,
we obtain
\be
\label{b4}
b_s(4)= \frac{b}{d}-  \frac{b^3}{2d^4}(46+97b^2+49 b^4),  
\en
\be
a_s(5)=-\frac{b^2}{d}-\frac{b^2}{2d^4}(25+49b^2+22 b^4),
\en
\be
a_s(7)=-\frac{b^2}{2d^2},
\en
\be
\label{b8}
b_s(8)=-\frac{b^3}{2d^2},
\en
where $b=b_s(2)$ and $d=5+7b^2$.
The magnitude of $b$ is determined by setting $\om=\sqrt{f_s(1)}$, 
Eq. \ref{om1}, and solving Eq. \ref{be} for $k_e=2$ which
takes the form  $b=h(b)$, where  
\be
\label{map1}
h(b)=\frac{g_s(2)}{4 f_s(1)}. 
\en
We find  that $h(b)$ is a positive monotonically decreasing function
for $b>0$ which diverges as $b$ approaches zero, and vanishes as 
$b$ increases indefinitely. Therefore, 
there exists  only a single solution which we find
at  $b=.3793$ corresponding to  $\om=1.448$.
This result  is in  good agreement with our numerical solution
of Eqs. \ref{ao} and  \ref {be}, which gives $b=.3776$ and $\om=1.452$.  
The resulting figure eight orbit obtained from the Fourier coefficients
given by  Eqs. \ref{b4} - \ref{map1} is shown in Fig. 3, 
which is graphically indistinguishable from  the orbit obtained 
from the numerical solution
of Eqs. \ref{ao} and  \ref{be}. Surprisingly, these analytic coefficients
also give a good approximation to the zero-angular-momentum 
figure eight orbit for $p < 2$, because it turns out that 
the moment of inertia is approximately constant
in this case \cite{simo}. As $p$ decreases 
we find that $b$ decreases and vanishes when $p=-2$, 
because in this limit of a harmonic force law 
the figure eight orbit becomes a linear collision orbit.  

\begin{figure}
\begin{center}
\epsfxsize=\columnwidth
\epsfbox{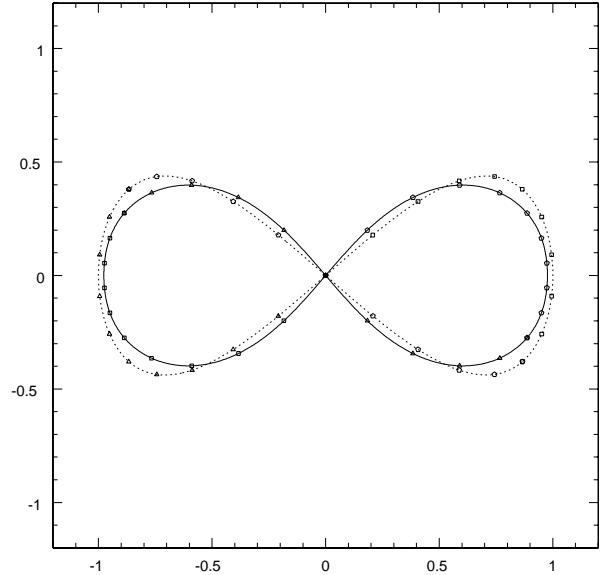}
\end{center}
\caption{
Two analytic approximations to the  figure eight periodic
orbit with zero angular momentum for $p=2$.
The dashed line is the solution for n=1,
a lemniscate, and the solid line is the solution
given by Eqs. 23-26 for n=4
}
\label{equi4}
\end{figure}

The zero-angular-momentum  orbit  provides the basis for a perturbation
calculation of the corresponding orbit for small values of the  
angular momentum.  Expanding the coefficients $a_c(k_e)$ and
$b_c(k_o)$  in powers of  $b_c(1)=b$, these coefficients  
together with the unperturbed coefficients
$a_s(k_o)$ and $b_s(k_e)$ determine the angular momentum of the orbit.
For example, for $p=1$ we find that $\Om = 1.647 b$, and 
substituting this value of $\Om$  and the lowest order  Fourier
coefficients in Eqs. \ref{ang2} and \ref{iner2} we obtain $ L_t=-.5276 b$.
Solutions for larger values of $b$  are obtained numerically
by iterating these equations, fixing  $a_s(1)$ and
starting with some arbitrary  value for  $b_s(2)$,  
while all other Fourier coefficients are initially set equal to zero. 
As $b$ increases the figure eight orbit becomes 
more  asymmetric with respect
to the y-axis, as expected from  the increasing  action of the centrifugal
and Coriolis forces in the rotating frame of reference, and 
at a critical value of the angular momentum 
the y-component of the  velocity of the particle at the end of the orbit
in a direction opposite to the rotation vanishes
giving rise to a  cusp. 
A further increase of the angular momentum then requires that this 
component of velocity change sign at this point, which  give
rise to an additional loop in the orbit, as shown in Fig. 4.

\begin{figure}
\begin{center}
\epsfxsize=\columnwidth
\epsfbox{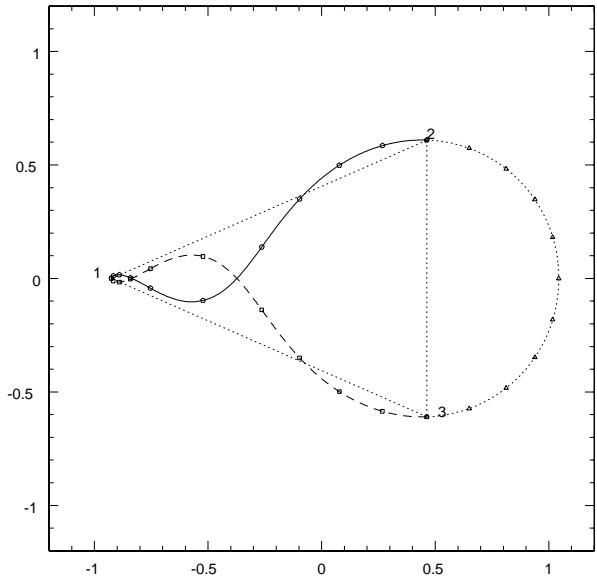}
\end{center}
\caption{
Asymmetric figure eight periodic orbit with finite angular
momentum in a rotating frame for $p=1$ and $b=.38$.
The additional loop shown at the left end
of the orbit forms when $ b>.34$.
}
\label{equi6}
\end{figure}

Finite angular momentum solutions  can also be obtained by rotating
the figure eight orbit about an axis in the plane of the original orbit
leading to a three dimensional orbit as shown in Fig. 5

\begin{figure}
\begin{center}
\epsfxsize=\columnwidth
\epsfbox{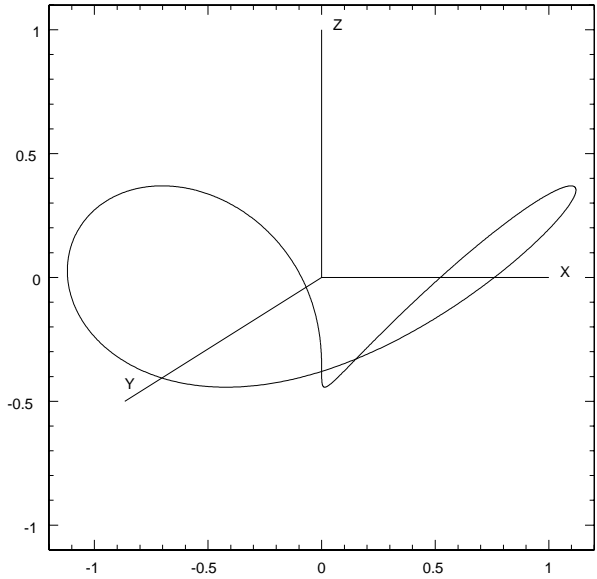}
\end{center}
\caption{
Three dimensional figure eight orbit with finite angular
momentum along the z-axis.
}
\label{equi7}
\end{figure}

Solutions of our equations provide initial conditions 
for a numerical integration of the equations of motion
which can then be applied  to study  the stability
of these  periodic orbits with respect to small perturbations.
A preliminary investigation indicates that some of these solutions are
stable over a restricted range of angular momentum and
values of $p$ \cite{stable1}.  Our method can readily be 
extended to investigate similar  periodic motions for an
arbitrary number of equal mass particles moving 
on the same orbit, many of which have  been discovered recently \cite{mont3}.
For example,  any odd number $n_o$ of particles can  travel
on a figure eight orbit similar to that shown in Fig. 3.  
We find that this orbit, Fig. 6, approaches 
a unique limit as $n_o\to \infty$, 
but the frequency increases logarithmically with $n_o$.

\begin{figure}
\begin{center}
\epsfxsize=\columnwidth
\epsfbox{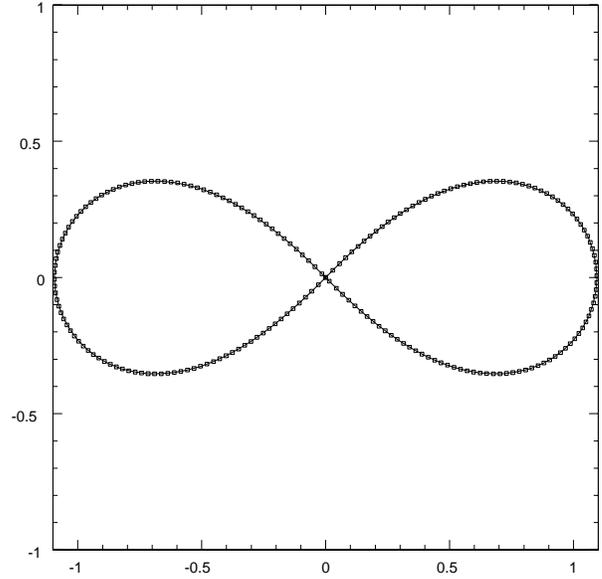}
\end{center}
\caption{
Figure eight orbit with 201 particles
}
\label{equi8}
\end{figure}

In contrast, for the extension of Lagrange's circular orbit 
each additional particle leads to a different orbital curve.   
Finally we have also applied our method to 
study the case of three particles with different masses  
which must move, however, along different orbital curves.
An example is shown in Fig. 7 which is an extension of 
an orbit discovered by Moore \cite{moore} 
for the case of equal mass particles.

\begin{figure}
\begin{center}
\epsfxsize=\columnwidth
\epsfbox{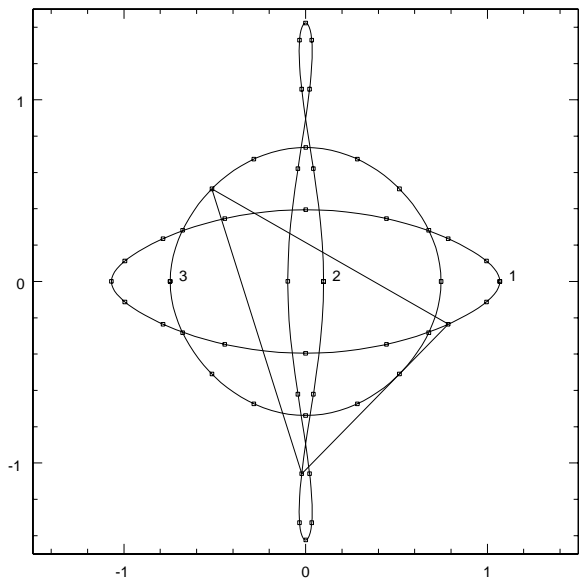}
\end{center}
\caption{
Periodic orbits for three  particles  with  masses 1.0,.5 and 1.5   
at initial positions labeled 1, 2, and 3 respectively,   
under the action of inverse square pairwise forces. Particles
1 and 2 move initially clockwise, 
while  particle 3 moves anti-clockwise.
Positions are shown at 1/16 of a period, and
the vertex of the  triangle shows their relative position after  1/8  
of a period. 
}
\label{equi9}
\end{figure}

\newpage

\subsection*{Acknowledgements}

I would like to thank Richard Montgomery for stimulating 
my interest in this problem and for many useful discussions,
and Alan Chenciner and Charles Sim\'o for  helpul comments.


\vskip -0.2in

\begin{thebibliography}{20}
\bibitem {moore}   C. Moore, Phys. Rev. Lett. 70 (1993) 3675.

\bibitem {mont1}   A. Chenciner and R.Montgomery, Annals of Mathematics 152 
                   (2000) 881.

\bibitem {mont2}   R. Montgomery, Notices of the American Mathematical 
                   Society (May,2001) 471.

\bibitem {simo}    C. Sim\'o, {\it Dynamical properties of the eight solution
                   of the three-body problem}, to appear in the Proceedings
                   of a Chicago conference dedicated to Don Saari (December
                   15-19, 1999). The numerical calculations in this paper
                   are an order of magnitude more precise than those 
                   of C. Moore, ref. 1, who used steepest descent 
                   on discretized orbits.

\bibitem{newton1}  Newton obtained the first periodic orbit for a 
                   three-body system in his treatment of the lunar problem, 
                   {\it Principia, Book 3, Prop. 28}. 
                   His solution  was later derived more fully  by L. Euler 
                   and by G.W. Hill.

\bibitem{lagrange} J.L. Lagrange, {\it Essai sur le Probl\'eme des Trois Corps},
                   Oeuvres (Paris 1873) Vol. 6, pp. 272-292.
	           Lagrange's solution can be readily extended to 
                   $n$ bodies on the vertices of an n-polygon.

\bibitem{plane}    Similar relations apply for rotation along any other
                   direction.


\bibitem{stable1}  According to the virial theorem, for $p \ge 2$ all orbits 
                   are unstable because the total energy $E \ge 0$.

\bibitem{mont3}   A.Chenciner,J. Gerver, R. Montgomery and C. Sim\'o,  
                  {\it Simple Choreographic Motions of N Bodies:
                  A preliminary Study}, (Montgomery,private communication)

\end{thebibliography}
\end{document}